\documentclass{article}

\usepackage{graphicx} 

\usepackage{lmodern}
\usepackage[top=1in, bottom=1in, left=1in, right=1in]{geometry}

\usepackage{amsmath,amssymb,amsthm}
\usepackage{wrapfig}
\usepackage{tikz}
\usetikzlibrary{positioning}
\usepackage{braket}

\newtheorem{theorem}{Theorem}

\newtheorem{definition}{Definition}

\newtheorem{result}{Result}
\newtheorem{question}{Question}
\newtheorem{lemma}[theorem]{Lemma}
\newtheorem{fact}[theorem]{Fact}

\newtheorem{proposition}[theorem]{Proposition}
\newtheorem{corollary}{Corollary}

\newtheorem*{lemma*}{Lemma}
\newtheorem*{proposition*}{Proposition}
\newtheorem*{corollary*}{Corollary}
\newtheorem*{theorem*}{Theorem}

\newcommand{\E}{\mathbb E}

\newcommand{\noisyR}{\operatorname{noisyR}}
\newcommand{\R}{\operatorname{R}}

\newcommand{\RS}{\operatorname{RS}}
\newcommand{\ACT}{\operatorname{ACT}}
\newcommand{\Ent}{\operatorname{Ent}}
\newcommand{\cost}{\operatorname{cost}}

\usepackage{xcolor}
\usepackage[normalem]{ulem} 

\usepackage[
    backend=biber,
    style=numeric-comp,
    sorting=nty,
    url=false,
    maxbibnames=99
]{biblatex}
\usepackage[
    colorlinks=true,
    allcolors=blue,
    citecolor=blue
]{hyperref}
\usepackage[capitalise,nameinlink]{cleveref}

\title{Composition Theorem for Randomized Query Complexity}

\author{%
\begin{tabular}{c@{\hspace{2.2em}}c}
Samruddhi Pednekar & Supartha Podder
\end{tabular}\\[0.9em]
{\small Stony Brook University, New York, USA}\\[0.4em]
{\small \texttt{\char`\{spednekar, supartha\char`\}@cs.stonybrook.edu}}
}

\date{}
\begin{document}

\maketitle

\begin{abstract}

For bounded-error randomized query complexity, we resolve the composition conjecture up to logarithmic factors. The general upper bound
$\R(f\circ g) = O\!\left(\R(f)\R(g)\log(\R(f)+1)\right)$ was already known. Although~\cite{meiburg2026perfectrandomizedcomposition} recently disproved constant-factor multiplicative composition for total Boolean functions, the conjecture allowing polylogarithmic losses remained open. We settle this question by proving that, for all (possibly partial) Boolean functions $f$ and $g$,
\[
\R(f\circ g)
=
\Omega\!\left(\frac{\R(f)\R(g)}{\log(n+1)}\right),
\]
where $n$ is the input size for $f$.

\end{abstract}

\tableofcontents

\newpage

\section{Introduction}
A longstanding problem in Boolean function complexity is to determine how the complexity of a block composition relates to the complexities of its constituent functions.
For total Boolean functions, tight composition theorems are known for deterministic query complexity and polynomial degree
\cite{savicky2002determinism, montanaro2013composition, tal_function_composition}, as well as bounded-error quantum
query complexity \cite{reichardt2011reflections, lee2011quantum}. Recently,~\cite{pednekar2026generalcompositiontheoremapproximate} proved a composition theorem for approximate degree and showed that constant-factor multiplicative composition fails for  exact quantum query complexity. In a latest development,~\cite{meiburg2026perfectrandomizedcomposition} showed that constant-factor multiplicative composition fails for randomized query complexity even for total Boolean functions. Nevertheless, composition for randomized query complexity up to polylogarithmic factors remained open
\cite{ben2020tight, ben2022randomised, chakraborty_et_al:LIPIcs.APPROX/RANDOM.2023.63}. In this work, we resolve this question by establishing a composition lower bound with a logarithmic loss for all (possibly partial) Boolean functions.

Decision trees have been a fundamental model in computational complexity and have been studied extensively in the form of deterministic, randomized and quantum settings. 
Randomized query complexity of a Boolean function $f$ is the minimum worst-case number of queries required by a randomized decision tree to compute $f$. A long-standing open question has been 
\begin{question}
    For all total Boolean functions $f,g$,
    \[\R(f \circ g) = \Omega(\R(f)\R(g))?\]
\end{question}
Recently,~\cite{meiburg2026perfectrandomizedcomposition} showed that this lower bound does not hold true by giving a counter-example. However, the same bound with polylogarithmic factors remained open for total Boolean functions.
\begin{question}\label{conjecture_poly_randomized}
    For all total Boolean functions $f,g$,
    \[\R(f \circ g) = \tilde{\Omega}(\R(f)\R(g))?\]
\end{question}

In the past, several papers have made progress towards answering whether composition of randomized query complexity holds by either showing lower bounds in terms of new related measures like noisy randomized query complexity, linearized randomized complexity, max-conflict complexity and randomized sabotage complexity or by showing the composition holds for some special class of families.~\cite{randomized_sabotage} introduced a measure called randomized sabotage complexity, denoted by $\RS(f)$ and showed that $\RS(f) = O(\R(f))$ and $\R(f \circ g) = \Omega(\R(f)\RS(g))$. They also asked if $\R(f) = \Theta(\RS(f))$ for all total Boolean functions which would resolve the open conjecture of composition of randomized query complexity. Later,~\cite{conflict_complexity} introduced a measure called max-conflict complexity, denoted by $\overline{\chi}$ and showed that $\overline{\chi}(g) \in \Omega(\sqrt{\R_{1/3}(g)})$ for every partial Boolean function $g$ and $\R_{1/3}(f \circ g) = \Omega(\R_{4/9}(f) \overline{\chi}(g))$ and $\overline{\chi}(g) \ge \RS(g)$. They also proved a relation $\R_{1/3}(f \circ g) = \Omega(\R_{4/9}(f)\sqrt{R_{1/3}(g)})$.~\cite{ben2020tight} introduced noisy oracle model and a measure called noisy randomized query complexity, denoted by $\noisyR$ and showed that for all Boolean functions $f$, $\noisyR(f) \le \R(f)$ and $\R(f \circ g) = \Omega(\noisyR(f) \R(g))$. They also gave a counter-example to $\noisyR(f) = \Omega(\R(f))$ and equivalently for $\R(f \circ g) = \Omega(\R(f)\R(g))$ using a partial function and thus refuted the constant-factor composition for partial Boolean functions. But the question remained open for the same bound with logarithmic factors. They conjectured that the following holds true 
\[ \noisyR(f) = \Omega\left(\frac{\R(f)}{\log{(n)}}\right). \]
Thus, this conjecture is equivalent to the following question
\begin{question}\label{conjecture_log_randomized}
    For all (possibly partial) Boolean functions $f,g$,
    \[\R(f \circ g) = \Omega\left(\frac{\R(f)\R(g)}{\log{(n)}}\right)?\]
    where $n$ is the input size of $f$.
\end{question}

Later,~\cite{chakraborty_et_al:LIPIcs.APPROX/RANDOM.2023.63} showed that $\noisyR(f)=\Theta(\R(f))$ for all Boolean functions with $\R(f)=\Theta(n)$. They also showed that if $\R$ composes with respect to an outer function, then $\noisyR$ also composes with respect to the same outer functions. Another paper~\cite{ben2022randomised} introduced a measure called linearized randomized complexity, denoted as $\mathsf{LR}$, which is the largest possible measure satisfying $\R(f \circ g) = \Omega(\R(f)\mathsf{LR}(g))$ for all partial Boolean functions $f,g$.

In this work, we show that bounded-error randomized query complexity composes multiplicatively with logarithmic factors. In particular, we establish that for every partial Boolean function, $\noisyR(f) = \Tilde{\Omega}(\R(f))$. The main result is stated below:

\begin{result}[Theorem~\ref{thm_noisy_randomized}, Corollary~\ref{cor_randomized_composition}]
    For a non-constant partial Boolean function $f: D \subseteq \{0,1\}^n \rightarrow \{0,1\}$, 
    \[ \R(f) \le C \log{(n+1)}\noisyR(f). \]
    Consequently, for all partial Boolean functions $f:D \subseteq \{0,1\}^n \rightarrow \{0,1\}$ and $g:D' \subseteq \{0,1\}^m \rightarrow \{0,1\}$,
    \[ \R(f \circ g) = \Omega\left( \frac{\R(f) \R(g)}{\log{(n+1)}}\right). \]
    
\end{result}

\paragraph{AI Disclosure.}
We used ChatGPT 6 Astra to explore research directions, identify potential errors in arguments, simplify proof strategies, and improve the exposition. Our use of AI involved sustained back-and-forth discussions over several weeks. We explored numerous approaches drawn from the literature, including recent work, repeatedly refined our arguments, and pursued both proofs and potential counterexamples. Our initial attempts to establish a randomized composition result were consistently unsuccessful. Following the appearance of recent work on randomized composition~\cite{meiburg2026perfectrandomizedcomposition}, we asked ChatGPT 6 Astra to prove a weaker bound allowing a logarithmic loss. It succeeded in deriving this bound, leading to the randomized composition result presented here. The authors have rewritten the proof and take full responsibility for the mathematical content and final presentation.

\paragraph{Acknowledgment.}
We thank Srinivasan Arunachalam for helpful discussions. 

\section{Technical Overview}

Noisy randomized model was introduced in~\cite{ben2020tight} and they showed that for all partial  Boolean functions $f$ and $g$,
\[\noisyR(f) \le \R(f), \quad\R(f \circ g) = \Omega(\noisyR(f)\R(g)).\] 
 The paper explicitly mentioned the following conjecture: for all (possibly partial) Boolean function~$f$, 
 \[\noisyR(f) = \Omega\left( \frac{\R(f)}{\log{(n+1)}}\right)?\] 
 
 Therefore, to prove the bound in Question~\ref{conjecture_poly_randomized}, it suffices to show that the above relation holds true.
This paper heavily builds on the concepts used in~\cite{minimax_randomized},~\cite{ben2020tight}. Let us first recap the noisy oracle model. Unlike the ordinary query model which reveals the queried bit exactly, in this noisy oracle model, a query with bias $\gamma$ returns the correct bit with probability $(1+ \gamma)/2$ and costs $\gamma^2$. The measure $\noisyR$ captures the query complexity of such model. Let $\mathcal{A}$ be an arbitrary noisy randomized decision tree that computes the given function $f$ with error at most $1/3$ on every promised input $x \in D$ where $D$ is the domain of $f$. Let $T$ be the stopping time of $\mathcal{A}$. At step $t \le T$, $\mathcal{A}$ adaptively chooses a coordinate $i_t$ and a query strength $\gamma_t \in [0,1]$. Since the cost of the query is $\gamma_t^2$, the total noisy cost of the execution is $\sum_{t=1}^T \gamma_t^2$. The cost of a noisy randomized decision tree on an input $x$ is given by $\operatorname{cost}(\mathcal{A},x) = \E\bigl[\sum_{t=1}^T \gamma_t^2\bigr]$ and then the cost of $\mathcal{A}$ is $\max_{x \in D}\{ \operatorname{cost}(\mathcal{A},x) \}$. Taking an infimum over the cost of all such noisy decision trees gives us $\noisyR(f)$. Thus, in order to prove the desired lower bound, we choose a hard distribution $\mu$ such that for every noisy decision tree $\mathcal{A}$, 
\[ \E_{X \sim \mu}\bigl[ \operatorname{cost}(\mathcal{A},X) \bigr] \ge \Omega\left(\frac{\R(f)}{\log{(n+1)}}\right). \]

This hard distribution $\mu$ is supplied by the minimax theorem of~\cite{minimax_randomized}. In particular, it is balanced, meaning $\Pr_{X \sim \mu}[f(X)=0] = \Pr_{X \sim \mu}[f(X)=1] =1/2$ and for every ordinary randomized algorithm $\mathcal{R}$ with terminal transcript $\mathsf{H}$ and number of exact queries $Q$,
\[ \E_{\mu}[\beta^2] \le \frac{c}{\R(f)}\E_{\mu}[Q], \]
where $\beta$ is the posterior label bias 
\[ \beta = \E[2f(X)-1 \; \mid \; \mathsf{H} ]. \]

We will keep a track of an arbitrary noisy decision tree at every posterior state after each query using a function that we call ``potential", labeled as $W$. Building on top of the forecasting algorithm given in~\cite{minimax_randomized}, we construct $W$ using auxiliary forecasting algorithms. Each such forecasting algorithm $\mathcal{B}$ gives a payoff which is the quantity below
\[ \text{Payoff}_{\mathcal{B}} =  \text{forecasting score} - \alpha \cdot \text{number of queries} + \text{entropy regularization}. \]

Entropy regularization is a standard way to replace a hard maximum by a smooth log-sum-exp expression~\cite{mdp_theory_geist}.
The potential is defined as the supremum of this expected payoff over all such auxiliary exact-query forecasting algorithms. After handling the case $R(f)\le 2c$ separately, we particularly set this query charge term to be $\frac{2c}{\R(f)}$ and the forecasting score to be the normalized Brier score.
Note that this algorithm $\mathcal{B}$ is neither the same as an ordinary bounded-error randomized algorithm nor the noisy algorithm $\mathcal{A}$ that computes $f$ with an error $1/3$. This is just an adaptive algorithm that may stop querying and output the best forecast from the information available to it. This algorithm follows a dynamic programming framework (see~\cite{dynamic_programming}), expressed by a Bellman recurrence: it may stop and give the optimal value or perform additional queries and incur the associated cost. 
 This potential, when evaluated at the algorithm's posterior state, is initially small under the chosen hard distribution $\mu$. The expected increase on each noisy query is bounded by $O(\alpha \log{(n+1)})$ times noisy query's cost and has a terminal expected value bounded below by a positive constant when the algorithm successfully terminates.

 Let us look at the initial bound. Given a transcript, the best expected score is the squared posterior bias. The chosen hard distribution bounds this squared posterior bias of every auxiliary algorithm in terms of its expected number of queries averaged over all transcripts. In particular, it gives an upper bound of $c/ \R(f)$ times expected query count. Choose a sufficiently large query charge, that is, $\alpha =2c/ \R(f) $ compared to this best expected score. Moreover, the entropy term is also sufficiently small. This makes the initial potential small. 
 \[ W(\mu) \le 1/100. \]
 
 Let us now bound the expected potential increase by one noisy query. When a noisy algorithm queries a bit, the observed answer re-weights the current posterior according to the query strength $\gamma$. Considering this re-weighted path, the expected increase in potential is a symmetric difference. For weak queries, meaning $\gamma \le 1/2$, the first-order terms cancel and hence the bound on second derivative gives the dependence on $\gamma^2$. For stronger queries, that is $\gamma > 1/2$, convexity and exact-query charge suffice.
 If $\gamma_t$ is the strength of the query $t$ and $\mu_{t-1}$ is the current posterior before the query, we get the following bound
\[ \E\left[W(\mu_t) \mid \mathsf{H}_{t-1}\right] - W(\mu_{t-1}) \le C_b' \alpha \log{(n+1)} \gamma_t^2. \]
Summing the one query bounds and telescoping over the finite computation gives us the following total bound.
  
\[ \E[W(\mu_T)] - W(\mu) \le C_b'\alpha \log{(n+1)}\E\left[\sum_{t=1}^T \gamma_t^2\right]. \]

Let us now consider the terminal potential. Since the noisy randomized algorithm computes $f$ within an error of $1/3$, its terminal posterior has expected squared bias at least $1/9$. Since the potential is at least the best score obtained by stopping immediately, its terminal expectation is also at least $1/9$. 
If $\mu_T$ is the posterior when the noisy algorithm $\mathcal{A}$ terminates, then using the success probability of $\mathcal{A}$, we get,
\[ \E[W(\mu_T)] \ge \frac{1}{9}. \]

This gives us a constant lower bound on the total potential increase. To achieve this, the expected total noisy cost has to be $\Omega(1/(\alpha \log{(n+1)}))$. Taking $\alpha = 2c/\R(f)$ gives us the desired bound.

\[ \E\left[\sum_{t=1}^T \gamma_t^2\right] = \Omega\left( \frac{1}{\alpha \log{(n+1)}} \right).  \]

Now let us look at the construction of $W$. An auxiliary forecasting algorithm $\mathcal{B}$ chooses a probability distribution $p_h$, depending on the history $h$, over the set of actions which is $\ACT = \{*,1,\dots,n\}$. If it chooses $*$, then it stops, otherwise for $j \in [n]$, it queries $x_j$ exactly and continues. We also use a separate entropy term which is the entropy of this action distribution multiplied by a sufficiently small constant times $\alpha/ \log{(n+1)}$. This keeps the entropy term at each decision small compared with the query charge.
For the forecasting score, normalized Brier score $\operatorname{score}_{Br}(y,z) = 1-4(y-z)^2$ is used. 

Let us now look at the motivation behind the entropy term in the payoff. Suppose a noisy algorithm queries coordinate $i$ with strength $\gamma$. Then by Bayes' rule and positive homogeneity of $W$, the expected potential after the query is 
\[  \frac{1}{2} W((1+\gamma (2x_i-1))\mu) + \frac{1}{2} W((1-\gamma (2x_i-1))\mu). \]

Taking $\Phi(s) = W((1+s (2x_i-1))\mu)$, the expected increase in potential is the symmetric difference $\frac{\Phi(\gamma)+\Phi(-\gamma)}{2} - \Phi(0)$. Since the first order derivatives cancel out because of symmetry, the second order derivative controls the bound. In order to evaluate that, we need $W$ to be sufficiently smooth in the relevant range and the entropy term ensures smoothness.

\section{Preliminaries}

\subsection{Definitions}

We use the standard block composition of Boolean functions, see~\cite{tal_function_composition}.
\begin{definition}[Composition of Boolean functions]
    Let $f: D'\subseteq \{0,1\}^n \rightarrow \{0,1\}$ and $g: D \subseteq \{0,1\}^m \rightarrow \{0,1\}$. Their block composition has domain 
    \[D_{f \circ g}= \left\{x=(x^{(1)},\dots,x^{(n)}) \in D^n: \left(g(x^{(1)}),\dots,g(x^{(n)})\right) \in D' \right\}. \]
    
    Then, for $x \in D_{f \circ g}$, define 
    \[ (f\circ g) (x) = f(g(x^{(1)}),\dots,g(x^{(n)})). \]
    $f$ is called the outer function and $g$ is called the inner function.
\end{definition}
We assume throughout that functions and compositions have nonempty promise domains. If either the outer function or the inner function is constant, then the randomized composition lower bound trivially holds.

\begin{definition}[Deterministic Decision Tree]
    A deterministic decision tree is a finite rooted binary tree whose internal nodes are labeled with index $i \in [n]$ and each leaf is labeled either $0$ or $1$.
    Each internal node has two outgoing edges corresponding to $0$ and $1$. For an input $x \in \{0,1\}^n$, it follows the path given by the coordinate $x_i$ and outputs the value given by the leaf. This deterministic decision tree is said to compute a Boolean function $f$ if it outputs $f(x)$ on each input $x$ in the domain.
    
\end{definition}

\begin{definition}[Randomized Query Complexity]
    A randomized decision tree $\mathcal{R}$ is a probability distribution over deterministic decision trees. Randomized query complexity of a Boolean function $f$, denoted as $\R_{\epsilon}(f)$, is the minimum worst case queries among all randomized decision trees computing $f$ with error at most $\epsilon$ on every input in the domain of $f$. When $\epsilon = 1/3$, we write $\R_{\epsilon}(f) =\R(f)$.
\end{definition}

\begin{definition}[Noisy Oracle Model~\cite{ben2020tight}]
    Let $x \in D \subseteq \{0,1\}^n$. A noisy query is specified by a coordinate $i \in [n]$ and has a strength $\gamma \in [0,1]$ associated with it. The oracle then returns $Y \in \{0,1\}$ satisfying 
    \[ \Pr[Y=x_i] = \frac{1+ \gamma}{2}, \quad \Pr[Y\ne x_i] = \frac{1- \gamma}{2}. \]
    The cost of this noisy query is $\gamma^2$.
    Every oracle query returns an independent random variable.
\end{definition}

\begin{definition}[Noisy Randomized Algorithm~\cite{ben2020tight}]

A noisy decision tree is a finite binary tree with internal nodes labeled as $(i,\gamma)$ where $\gamma \in [0,1]$ and $i \in [n]$ and each leaf is labeled $0$ or $1$. Descendants can have the same node label as their ancestors. Let $\mathcal{A}$ be a noisy oracle randomized algorithm. Then it is a finite support distribution over finite noisy decision trees. A noisy randomized query algorithm might adaptively choose the coordinate $i_t$ and query strength $\gamma_t \in [0,1]$ for its $t$-th query as a function of its previous queries, answers and internal randomness. The cost of execution is $\sum_{t=1}^T \gamma_t^2$ where $T$ is the stopping time of $\mathcal{A}$. 

Let $f: D \subseteq \{0,1\}^n \rightarrow \{0,1\}$. It is said to compute a function $f$ with error up to $\epsilon$ if $\Pr[\mathcal{A}(x)\ne f(x) ] \le \epsilon$ for all $x \in D$. For a fixed input $x \in D$, define the cost of $\mathcal{A}$ on input $x$ as 
\[ \operatorname{cost}(\mathcal{A},x) = \E\left[\sum_{t=1}^T \gamma_t^2\right], \]
where the expectation is taken over the internal randomness of $\mathcal{A}$ and the noisy oracle answers when $\mathcal{A}$ is run on $x$.
\end{definition}

\begin{definition}[Noisy Randomized Query Complexity~\cite{ben2020tight}]
The noisy randomized query complexity of $f: D \rightarrow\{0,1\}$ is 
\[ \noisyR_{\epsilon}(f) = \inf_{\mathcal{A}} \max_{x \in D}\{ \operatorname{cost}(\mathcal{A},x) \}, \]
where infimum is taken over all noisy randomized query algorithms computing $f$ with error at most $\epsilon$. 

When $\epsilon = 1/3$, the subscript $\epsilon$ is dropped.
\end{definition}

\begin{definition}[Forecasting Algorithm~\cite{minimax_randomized}]
    Forecasting algorithms are generalized randomized algorithms that output a confidence value. Unlike ordinary randomized algorithms that output either $0$ or $1$, forecasting algorithms output a value in $[0,1]$ for $f(x)$ for an input $x$. The confidence value is considered to be low if it is close to $1/2$, whereas, it is considered high if it is close to $0$ or $1$.
\end{definition}

Following the definition of transcript from~\cite{ben2020tight}, our complete transcript includes the tree sampled independently of the input, along with the queries and observed answers, including the query strengths in the noisy model. $\mathsf{H}_t$ denotes the transcript after $t$ queries. 

\begin{definition}[Posterior Label Bias]
Let $f: D \subseteq \{0,1\}^n \rightarrow \{0,1\}$ be a partial Boolean function. Let $X \sim \mu$ for some probability distribution $\mu$ supported on $D$. Let $\mathsf{H}$ be the transcript of a query algorithm run on $X$. Then,
the corresponding posterior label bias, denoted as $\beta_{\mathsf{H}}$ is the value $\E[2f(X)-1 \mid \mathsf{H}]$.
    Note that $\beta_{\mathsf{H}}$ is a random variable.
\end{definition}

\begin{definition}[Entropy] We use the standard discrete Shannon Entropy as the entropy defined below
   \[\Ent(p)  =\sum_{a : p_a >0} p_a \log{\frac{1}{p_a}},\] 
\end{definition}
with the convention $0 \log 0 =0$ and natural logarithms.

\subsection{Facts and Previous Known Results}
\begin{lemma}[Lower bound on noisy randomized query complexity~\cite{ben2020tight}]\label{lem_noisy_constant_lower}

For a non-constant partial Boolean function $f$,
\[ \noisyR(f) = \Omega(1). \]
\end{lemma}

\begin{lemma}[Lower Bound on Block Composition~\cite{ben2020tight}]\label{lem_noisy_lower_bound}
    For all partial Boolean functions $f,g$, 
    \[\R(f \circ g) = \Omega(\noisyR(f)\R(g)).\]
\end{lemma}
The following Lemma is a consequence of a result from~\cite{minimax_randomized}.
\begin{lemma}[Existence of Hard distribution for forecasting~\cite{minimax_randomized}]\label{lem_hard_dist_existence}
Let $f: D \subseteq \{0,1\}^n \rightarrow \{0,1\}$ be a non-constant partial Boolean function.
There exists a balanced distribution $\mu$ on $D$ such that, for $X \sim \mu$, every ordinary randomized tree with complete terminal transcript $\mathsf{H}$ and query count $Q$ satisfies
\[ \E_{\mu}[\E(2f(X)-1 \mid \mathsf{H})^2] \le \frac{c}{\R(f)}\E_{\mu}(Q),\]
for an absolute constant $c\ge 1$ independent of $f$ and the tree. The expectation includes the tree's internal randomness. 
\end{lemma}
\begin{proof}
    Choose $\mu_0,\mu_1$ supported on $f^{-1}(0),f^{-1}(1)$ respectively as guaranteed by~\cite{minimax_randomized} and set $\mu = \frac{\mu_0+\mu_1}{2}$. Let $P_0,P_1$ be the transcript distributions under $\mu_0,\mu_1$ respectively.
    Under the balanced distribution, $\E_{\mu}[\E(2f(X)-1 \mid \mathsf{H})^2]$ is the symmetrized $\chi^2$ distance between $P_0$ and $P_1$ (see~\cite{minimax_randomized} Definition 3.5). If $h^2(P_0,P_1)$ is the squared Hellinger distance between them, then the symmetrized $\chi^2$ distance is at most $2h^2(P_0,P_1)$ using Lemma 3.6 from~\cite{minimax_randomized}. From Theorem 4.5~\cite{minimax_randomized} we get $h^2(P_0,P_1) \le 240 \E[Q]/\R(f)$, and taking $c=480$, we get the desired bound.
\end{proof}

\begin{fact}[Entropy Bound]\label{fact_entropy_bound}
    Let $p$ be a probability distribution on the $N$ available actions from $\ACT$. Then \[\Ent(p) \le \log{N}.\] 
\end{fact}
\begin{proof}
    Applying Jensen's inequality since $\log$ is concave, we get 
    \[ \Ent(p) = \sum_{a : p_a >0} p_a \log{\frac{1}{p_a}} \le \log\left(\sum_{a:p_a > 0} p_a \frac{1}{p_a} \right) = \log{|\operatorname{supp}(p)|} \le \log N. \]
\end{proof}

We get the following equality using the standard log-sum-exp identity. See~\cite{boyd2004convex} example 3.25. We adapt it for our notation.
\begin{fact}\label{fact_log_exp_sum}
    Let $\eta >0$ and let $p$ be a vector such that $p_a \ge 0$ for $a \in \ACT$ and $\sum_{a \in \ACT} p_a = 1$. Let $q_a \in \mathbb{R}$ for $a \in \ACT$. Then,
    \[ \max_{p} \sum_{a \in \ACT} p_a q_a + \eta \Ent(p) = \eta \log\left[ \sum_{a \in \ACT} e^{q_a/\eta} \right]. \]
    All logarithms are natural.
\end{fact}

\begin{fact}\label{fact_convex_upper_bound}
    Let $F$ be a convex function. For $x \in [-1,1]$,
    \[ F(x)+ F(-x) \le F(1)+ F(-1). \]
\end{fact}
\begin{proof}
    We can write $x =\frac{(1+x)}{2} 1  + \frac{(1-x)}{2} (-1) $ and similarly $-x =\frac{(1-x)}{2} 1  + \frac{(1+x)}{2} (-1)$. By convexity, $F(x) \le \frac{(1+x)}{2} F(1) + \frac{(1-x)}{2} F(-1)$ and $F(-x) \le\frac{(1-x)}{2} F(1)  + \frac{(1+x)}{2} F(-1)$. Adding these two equations gives us $F(x)+F(-x) \le F(1)+F(-1)$
\end{proof}

\begin{fact}\label{fact_symmetric_difference}
    Let $F$ be a twice continuously differentiable function on $[-a,a]$ for $a >0$. If $F''(x) \le M$ for all $x \in [-a,a]$, then for every $x \in [-a,a]$,
    \[ F(x)+F(-x) \le 2F(0)+Mx^2. \]
\end{fact}
\begin{proof}
    Without loss of generality, assume $x \ge 0$.
    Using Taylor's theorem with integral remainder, we obtain 
    \[ F(x) =F(0) + x F'(0) + \int_{0}^{x} (x-s ) F''(s) ds.\]
    Using $F''(s) \le M$ when $s \in [-a,a]$ and $x-s \ge 0$ when $s \in (0,x)$, the integral term is bounded above by $M(x^2/2)$ and hence $F(x) \le F(0)+xF'(0) +M(x^2/2) $.
    Similarly, $F(-x) \le F(0)- x F'(0)+M\left( \frac{x^2}{2}\right)$ and therefore adding $F(x)$ and $F(-x)$, we get
    \[ F(x)+F(-x) \le 2F(0) +M x^2. \]
\end{proof}

\section{Main Results}

\subsection{Constructing the Potential and auxiliaries}

In this subsection we first construct the potential $W$ that will be tracked at each query. This potential will be used throughout in the rest of the paper.

We will consider a finite randomized forecasting algorithm $\mathcal{B}$ with number of queries $Q_{\mathcal{B}}$. We define the action set for this algorithm as 
\[ \ACT = \{*,1,\dots,n\}, \]

where $*$ denotes stopping.

For a label $y \in \{0,1\}$ and forecast $z \in [0,1]$, we use normalized Brier score, denoted henceforth as $\operatorname{score}_{Br}$:
\[  \operatorname{score}_{Br}(y,z) = 1-4(y-z)^2. \]

Note that Brier score is already used in literature and~\cite{minimax_randomized} also uses it.

For a sufficiently large constant $b \ge 100$, fix parameters 
\[ \frac{1}{n} \le \alpha \le 1, \quad N = n+1, \quad \tau = \frac{\alpha}{b \log{N}}.  \]

After $t$ queries, the algorithm chooses its next action according to a distribution $p_t$ at the visited history over $\ACT$. The distribution $p_t$ is specified before drawing the next action. The algorithm could repeat queries to the same coordinates. Also the entropy sum includes stopping decision. Let $Z \in [0,1]$ denote the forecast of $\mathcal{B}$.
For each $x \in D$, define the expected payoff of $\mathcal{B}$ as
\[ L_{\mathcal{B}}(x) = \E\left[ \operatorname{score}_{Br}\bigl(f(x),Z\bigr) - \alpha Q_{\mathcal{B}} + \tau \sum_{t=0}^{Q_{\mathcal{B}}} \Ent(p_t) \right], \]

where the expectation is taken over random action choices of $\mathcal{B}$ run on input $x$.

    Let $\nu(x)$ be a non-negative measure on $D$. Define mass of $\nu$ as $m = m(\nu) = \sum_x \nu(x)$ and the stopping value as 
    $S(0)=0$ and for $\nu \ne 0$,
     \begin{equation}\label{eq_stopping_value}
     S(\nu) =  \sup_{z \in [0,1]}\sum_x \nu(x)\operatorname{score}_{Br}(f(x),z)=  m(\nu) \left(\frac{\sum_x \nu(x)(2f(x)-1)}{m(\nu)} \right)^2.    
     \end{equation}
     The equality in the above equation occurs by taking the maximum which occurs at $z = \frac{\sum_x \nu(x)f(x)}{m(\nu)}$.
    When the context is clear, shorthand $m$ will be used for $m(\nu)$.

    When $\nu$ is a probability distribution, 
    \[ S(\nu) = \bigl(\E_{X \sim \nu}[(2f(X)-1)]\bigr)^2. \]
    Furthermore, if $\mu_t$ is a posterior distribution $\mu_t(x) = \Pr[X=x \mid \mathsf{H}_t]$ after $t$ queries, by taking $\beta_t$ as the posterior label bias, we get $S(\mu_t) = \beta_t^2$.

    Also, note that $S$ is convex.
\begin{definition}[Potential]
    For a non-negative measure $\nu$, we now define the potential as
    \begin{equation}\label{eq_potential_definition}
        W(\nu) = \sup_{\mathcal{B}} \sum_{x \in D} \nu(x) L_{\mathcal{B}}(x).     
    \end{equation}
\end{definition}
 
    Let $\nu_j^{\sigma}$ for $\sigma \in \{0,1\}$ be the restriction of $\nu$ to $x_j$: 
    \[ \nu_j^1(x) = \nu(x)\mathbf{1}_{x_j=1} , \quad \nu_j^0(x) = \nu(x)\mathbf{1}_{x_j=0}. \]

    \begin{definition}[Query Actions]
        For a measure $\nu$ of positive mass $m=m(\nu)$,  define $q_a$ to be the supremum of expected payoffs obtained after fixing the next action to be $a \in \ACT$. This does not include the entropy term associated with the next action distribution.

    The value of stopping is 
    \[  q_*(\nu) = S(\nu). \]
    For every $j \in [n]$, the value of querying coordinate $j$ is 
    \begin{equation}\label{eq_query_actions}
         \qquad q_j(\nu)= -\alpha m+W(\nu_j^1)+W(\nu_j^{0}),  
    \end{equation}    
    where the current query is charged $-\alpha m$ and $W(\nu_j^{1}),W(\nu_j^{0})$ include subsequent query charges and entropy terms depending on the coordinate $j$.
    \end{definition}
    
    We now state some properties of this potential below:
    \begin{lemma}[Properties of the potential]\label{lem_basic_potential_properties}
    The potential $W$ given in Equation~\ref{eq_potential_definition} satisfies the following properties:
        \begin{enumerate}
            \item[(i)] W is convex and positive homogeneous and lower semicontinuous. 
            
            \item[(ii)]For every non-negative measure $\nu$,  
            \begin{equation}\label{eq_W_S}
                S(\nu) \le W(\nu) \le \left(1+ \frac{\alpha}{b}\right) m(\nu). 
            \end{equation}
            \item[(iii)] For a non-negative measure $\nu$ with $m(\nu)>0$,  \begin{equation}\label{eq_exp_sum_W}
         W(\nu) = m \tau \log{\sum_{a \in \ACT} e^{q_a(\nu)/(m \tau)} }.
        \end{equation}
        The maximizing distribution in the first-action optimization is
        \[ \pi_a = \exp\left( \frac{q_a(\nu)-W(\nu)}{m \tau} \right). \]
         Moreover, every query action satisfies the bounds
         \begin{equation}
             W(\nu) - \alpha m \le q_j(\nu) \le W(\nu).
         \end{equation}

            \item[(iv)] Fix a probability distribution $\nu$ and coordinate $i$. For $\chi_i(x) = 2x_i - 1$ , the map $ s \longmapsto W((1+s \chi_i)\nu)$ is smooth for $s \in (-1,1)$ and continuous for $s \in [-1,1]$.  
        \end{enumerate}

    \end{lemma}
    \begin{proof}
    \begin{enumerate}
        \item [(i)] Positive homogeneity of $W$ is immediate from its definition. For a fixed auxiliary forecasting algorithm $\mathcal{B}$, the map $ \nu \longmapsto \sum_x \nu(x) L_{\mathcal{B}}(x)$ is a continuous linear function in its input measure $\nu$. Taking a supremum over these linear functions makes $W$ convex and lower semicontinuous. 
    \item[(ii)] Since $W$ is an optimal value, stopping immediately instead of continuing to query gives us $W(\nu) \ge S(\nu)$. 
    Now, let us look at the entropy term in the payoff. First suppose that $\nu$ is a probability distribution. Since the normalized Brier score can be at most 1, on every execution of auxiliary algorithm making $Q$ queries, the payoff is at most $1 - \alpha Q + \tau \sum_{t=0}^Q\Ent(p_t)$ where $p_t$ is the action distribution at query $t$. Using Fact~\ref{fact_entropy_bound}, we get each payoff is at most
    \begin{equation}\label{eq_payoff_bound}
     1 - \alpha Q + \tau (Q+1) \log{N} = 1- \alpha Q + (Q+1) \frac{\alpha}{b} = 1 +\left(\frac{\alpha}{b} -\alpha\right)Q + \frac{\alpha}{b} \le 1+ \frac{\alpha}{b}.     
    \end{equation}
    
    Taking expectation and supremum over all such auxiliary algorithms, we get $W(\nu) \le  1+ \frac{\alpha}{b}$. 

    Now, let $\nu'$ be an arbitrary non-negative measure. When $\nu'=0$, $W(\nu')=0$, otherwise, by positive homogeneity of $W$, we get
    \[ W(\nu') = m(\nu') W\left(\frac{\nu'}{m(\nu')}\right) \le m(\nu') \left(1+ \frac{\alpha}{b}\right). \]
    \item [(iii)]
        
        Fix a first-action probability distribution $p$. Therefore, considering the entropy term, the best possible value is $\sum_a p_a q_a(\nu)+m \tau \Ent(p)$. Thus the potential $W$ is 
        \[ \sup_{p} \sum_a p_a q_a(\nu)+m \tau \Ent(p). \]

        Using standard log-sum-exp identity Fact~\ref{fact_log_exp_sum} with $\eta=m \tau$ and $q_a = q_a(\nu)$, we get,
        \[ \sup_{p} \sum_a p_a q_a(\nu)+m \tau \Ent(p) = m \tau \log\left(\sum_{a \in \ACT} e^{\frac{q_a(\nu)}{m \tau}}\right) \]
        If $\pi_a$ is the maximizing distribution, we get, 
        \[ W(\nu) = m \tau \log\left(\sum_{a \in \ACT} e^{\frac{q_a(\nu)}{m \tau}} \right), \quad \pi_a = \exp{\left(\frac{q_a(\nu) - W(\nu)}{m \tau}\right)}. \]

        These equalities hold even if an optimal tree does not exist. Since there are finitely many first actions, we can use arbitrarily close finite continuation for each answer.
        
        Convexity and positive homogeneity of $W$ imply subadditivity and therefore, we get $W(\nu) \le W(\nu_j^0)+W(\nu_j^1)$ and since $q_j(\nu)+ \alpha m = W(\nu_j^0)+W(\nu_j^1)$, we get,
        \begin{equation}
            W(\nu) - m \alpha \le q_j(\nu).     
        \end{equation}

        Additionally, since for every $a \in \ACT$, $\log\left(\sum_{a \in \ACT} e^{\frac{q_a(\nu)}{m \tau}} \right) \ge \frac{q_a(\nu)}{m \tau}$, we get
        \[  W(\nu) \ge q_a(\nu).  \]
        Let $\mathcal{J}$ be the set of determined coordinates and $|\mathcal{J}|=k$. Let $\ACT^r = \{*\} \cup ([n]\backslash \mathcal{J})$.
        Now, let $j \in [n]$ be constant on $\operatorname{supp}(\nu)$. Without loss of generality, let $x_j=1$ for every $x \in \operatorname{supp}(\nu)$. Then $\nu_j^0=0$ and $\nu_j^1=\nu$. Since $W(0)=0$, we get 
        \begin{equation}\label{eq_determined_coordinate_W}
            q_j(\nu) = -\alpha m+ W(\nu)+W(0) = W(\nu)-\alpha m.    
        \end{equation}
         Hence, for every determined coordinate, $e^{q_j(\nu)/{m \tau}} = e^{W(\nu)/{m \tau}}e^{-\alpha/\tau}$. 
        Since there are $k$ determined coordinates, exponentiating Equation~\ref{eq_exp_sum_W} gives 
        \[ e^{W(\nu)/(m \tau)} = e^{S(\nu)/(m \tau)}+ \sum_{j \in ([n]\backslash \mathcal{J})} e^{q_j(\nu)/(m \tau)}+ ke^{W(\nu)/{m \tau}}e^{-\alpha/\tau}.  \]
        Rearranging, taking $\log$ and multiplying by $m \tau$ gives us
        \begin{equation}\label{eq_undetermined_W}
            W(\nu) = m \tau \log{\left( \frac{e^{S(\nu)/(m \tau)}+ \sum_{j \in ([n]\backslash \mathcal{J})} e^{q_j(\nu)/(m \tau)}}{1-ke^{-\alpha/\tau}}\right)}
        \end{equation}

        Finally, we have $e^{-\alpha/ \tau} = e^{-b \log{N}} = N^{-b}$. Since $k \le n < N^b$, we get that the denominator in above equation is positive.
    \item [(iv)] 
    Smoothness is proved by using induction on the support size where the support is non-empty. On singleton support, $W$ is linear in $m$ by positive homogeneity. For a larger fixed support, every undetermined query has children with strictly smaller supports and by induction hypothesis, $W$ is smooth on these child measures. Smoothness of $S$ and the fact that denominator $1-kN^{-b}$ is a positive constant on respective support, Equation~\ref{eq_undetermined_W} makes $W$ smooth there.
    We know that the map is convex and finite and hence it is continuous in the interior, that is $(-1,1)$. Let us look at the continuity at the end-points.
    Since $W$ is lower semicontinuous, along any coordinate reweighting path $\nu_s = ((1+ \chi_i(x) s) \nu)$, 
    \[ W(\nu_1) \le \liminf_{s \to 1} W(\nu_s), \quad W(\nu_{-1}) \le \liminf_{s \to -1} W(\nu_s). \]
    Moreover, using convexity of this map, we get
    \[W(\nu_1) \ge \limsup_{s \to 1} W(\nu_s), \quad W(\nu_{-1}) \ge \limsup_{s \to -1} W(\nu_s).\]
    Thus, we get continuity at the endpoints $s = 1,s=-1$.
    For $|s| < 1$, the measure $(1+s \chi_i)\nu$ has the same support as $\nu$ and the map $s \longmapsto W((1+s \chi_i(x))\nu)$ is therefore smooth on $s \in (-1,1)$. 
    Thus, the map is continuous in $[-1,1]$ and smooth in $(-1,1)$.
    \end{enumerate}
        
    \end{proof}

\subsection{Curvature Bound}

A noisy query of coordinate $x_i$ reweights the current posterior $\nu$ by a factor of $1\pm \gamma$ depending on the observed answer. Denoting $s=\pm \gamma$, the path $(1+ s \chi_i)\nu$ interpolates between the current state at $s=0$ and the two exact query outcomes at $s=\pm 1$.

Throughout this subsection we will use fixed notation. In this subsection, we evaluate the bound for second order derivative of $W$ along a coordinate re-weighted path $(1+s \chi_i)\nu$, where $\chi_i(x)=2x_i-1$. When $s=0$, it represents the current state $\nu$. 
\[ \nu_s = (1+s \chi_i)\nu. \]
Fix a probability distribution $\nu$ and a coordinate $i$ and for $s \in [-1,1]$, define 
\[ \Phi(s) = W((1+s\chi_i)\nu), \quad m_s = 1+s\E_{\nu}[\chi_i(X)]. \]

We need to quantify the variation of $\Phi$ at its endpoints with its tangent line at $s=0$.

\begin{equation}\label{eq_D_pm}
    D^+ = \Phi(1) - \Phi(0) - \Phi'(0), \qquad D^- = \Phi(-1) - \Phi(0) + \Phi'(0).
\end{equation}

Similarly, $q_a(s)=q_a(\nu_s)$ and we need analogous comparisons for $q_a$.
\begin{equation}\label{eq_q_a_pm}
    d^+_a = q_a(1) - q_a(0) - q_a'(0), \qquad d_a^- = q_a(-1) - q_a(0) + q_a'(0).
\end{equation}

For $u,v \ge 0$, write 
\begin{equation}\label{eq_harmonic}
    H(u,v) = \frac{uv}{u+v}, \quad H(0,0)=0.
\end{equation}
    Note that $H$ is homogeneous. Additionally, $H(u,v)$ can also be written as $H(u,v) = \min_z \{z^2u+(1-z)^2v\}$ and thus, it is concave. By taking $z = 1/2$, we get $H(u,v) \le \frac{u+v}{4}$.

    Concavity and homogeneity imply superadditivity and thus we get 
    \begin{equation}\label{eq_superadditivity_H}
       H(u_1,v_1) + H(u_2,v_2) \le H(u_1+u_2,v_1+v_2) .
    \end{equation}

    Since $\Phi(s)=W((1+s \chi_i)\nu)$, we need to bound $\Phi''(0)$.

    \[\Phi(s) =  m_s\tau \log{\sum_{a \in \operatorname{ACT}}{e^{q_a(s)/(m_s\tau)}}}, \quad m_s > 0. \]

\begin{proposition}[Coordinate curvature bound]\label{prop_curvature_bound}
For every probability distribution $\mu$ and every coordinate $i \in [n]$,
    \begin{equation}
        \left. \frac{d^2}{ds^2} W((1+s \chi_i)\mu) \right|_{s=0} \le C_b \alpha \log{N},
    \end{equation}
    where $C_b$ is a constant depending on $b$.
\end{proposition}

Proofs for Proposition~\ref{prop_curvature_bound} and Lemma~\ref{lem_softmax_bound} are deferred to Appendix~\ref{appendix_softmax_bound}.

Define $\pi_a = \exp\left( \frac{q_a(0)-\Phi(0)}{\tau} \right)$. The expectation in the following Lemma is taken under $\pi_a$.

\begin{lemma}\label{lem_softmax_bound}
    Fix a coordinate $i$ and a distribution $\nu$ and assume $i$ is undetermined on $\operatorname{supp}(\nu)$. For $s \in [-1,1]$, write $\Phi(s)= W\left( (1+ \chi_i(x) s)\nu \right)$. 

   For $K= 36 b \log{N}$ and $\varepsilon = \frac{\alpha}{N^4}$,
    \[ \Phi''(0) \le \E[q_a''(0)]+K\left( H(D^+,D^-) - \E[H(d_a^+,d_a^-)] \right) +\varepsilon. \]
\end{lemma}

\begin{proposition}\label{prop_uniform_curvature_bound}
    For a probability distribution $\mu$ and $i \in [n]$, and $\chi_i(x) = 2 x_i -1$ for $x \in D$, let 
    \[ \Phi(s) = W((1+ \chi_i s)\mu). \]
    Then, for every $s \in [-1/2,1/2]$, 
    \[ \Phi''(s) \le \widehat{C}_b \alpha \log N,\]
    where $\widehat{C}_b$ is a constant that depends on $b$. 
\end{proposition}
\begin{proof}
    Let $r = \E_{\mu}[\chi_i(X)]$ and $m_s = 1+rs$. Also define 
    \[\mu_s = \frac{(1+s \chi_i)\mu}{m_s}, \quad F_s(t)=W((1+t \chi_i)\mu_s).\]
    Note that $(1+s \chi_i)\mu$ is a non-negative measure for $s \in [-1/2,1/2]$ since $\chi_i(x) \in \{-1,1\}$ and thus, $(1+s \chi_i(x)) \ge 1-|s| \ge 1/2$. Moreover, its mass is equal to $\sum_x (1+ \chi_i(x)s)\mu(x) =\sum_x \mu(x) +\sum_x \mu(x) \chi_i(x)s  = 1+\E_{\mu}[\chi_i(X)]s = 1+rs = m_s$. Thus, dividing by its mass gives us that it is a probability distribution.
    For a sufficiently small $h$, since $\chi_i^2(x)=1$, we get
    \[ \Phi(s+h) = m_s\left( 1- \frac{sh}{1-s^2}\right)F_s\left( \frac{h}{1-s^2-sh} \right). \]

    Differentiating twice at $h=0$ gives us 

    \[ \Phi''(s) = \frac{m_s}{(1-s^2)^2}F''_s(0).\]

    Applying Proposition~\ref{prop_curvature_bound}, we get $F_s''(0) \le C_b \alpha \log{N}$ for some constant $C_b$ depending on $b$. For $|s| \le 1/2$, $m_s \le 3/2$ and $(1-s^2) \ge 3/4$, we get $ \Phi''(s) \le \frac{8}{3} C_b \alpha \log N$. Taking $\widehat{C}_b = \frac{8}{3} C_b$ gives us the required bound.
\end{proof}
\subsection{Controlling increase in Potential}

\begin{proposition}[One query potential increase]\label{prop_one_potential_increase}
    Suppose $\mathcal{A}$ is a noisy randomized decision tree and let $\gamma \in [0,1]$ be the strength of one noisy query.
    Let $\mu$ be a probability distribution on $D$. Let $Y \in \{0,1\}$ denote the answer and let $\mu_y(x) := \Pr[X=x \mid Y=y]$ be the posterior corresponding to answer $y$. Then 
    \[ \E_Y[W(\mu_Y)] - W(\mu) \le C_b' \alpha (\log{N}) \gamma^2. \]
\end{proposition}
\begin{proof}
    Let $\mu$ be the current posterior distribution. $\mu,i,\gamma$ are fixed conditioned on the transcript before the query.
    Suppose a noisy query with strength $\gamma$ is made on coordinate $i$. Let $p_y = \Pr[Y=y] $ denote the probability of answer $y$ and for $p_y>0$, $\mu_y$ denote the corresponding posterior (with $p_y W(\mu_y)= 0$ when $p_y=0$).
    We use the re-centered coordinates $\chi_i(x) = 2 x_i -1$.
    By the definition of a noisy query, we get $\Pr[Y=y \mid X=x] =\frac{1+\gamma (2y-1) \chi_i(x)}{2}$. Using Bayes' rule, we get 
    \begin{equation}\label{eq_bayes_rule}
       p_y \mu_y(x) = \frac{1+\gamma (2y-1) \chi_i(x)}{2} \mu(x). 
    \end{equation}
    
    We can now write $\E_Y[W(\mu_Y)] = \sum_{y \in \{0,1\}} p_yW(\mu_y)$. By positive homogeneity of $W$, we can write $p_y W(\mu_y) = W(p_y \mu_y)$ and hence $\E_Y[W(\mu_Y)] = \sum_{y \in \{0,1\}} W(p_y \mu_y)$. Substituting Equation~\ref{eq_bayes_rule}, we obtain 
    \[\E_Y[W(\mu_Y)] = \frac{1}{2}W\left((1+\gamma \chi_i)\mu\right)+\frac{1}{2}W\left((1-\gamma \chi_i)\mu\right).\]

    Taking $\Phi(s) = W((1+s \chi_i)\mu)$ for $s \in [-1,1]$, we get
    \[  \E_Y[W(\mu_Y)] = \frac{\Phi(\gamma) + \Phi(-\gamma) }{2}. \]

    Let $ \Delta_{\gamma} := \E_Y[W(\mu_Y)] - W(\mu)$ and hence $\Delta_{\gamma} = \frac{\Phi(\gamma) + \Phi(-\gamma)}{2} - \Phi(0)$. It suffices to bound $\Delta_{\gamma}$ from above with the desired bound.
    
    $\Phi$ might not be smooth at the endpoints $+1,-1$. 
    We treat separately the two cases $\gamma \le 1/2$ and $\gamma>1/2$. For the case $\gamma \le 1/2$, Proposition~\ref{prop_uniform_curvature_bound} can be applied and for the case $\gamma>1/2$, a coarse bound can be obtained using convexity and that suffices.

    \textbf{Case 1}: $\gamma \le 1/2$.
    By Proposition~\ref{prop_uniform_curvature_bound}, we have $\Phi''(s) \le \widehat{C}_b \alpha \log N$ for $s \in [-1/2,1/2]$.
    Using Fact~\ref{fact_symmetric_difference}, we get 
     \[\Delta_{\gamma} \le \frac{\widehat{C}_b}{2} \alpha (\log N) \gamma^2.\]

    \textbf{Case 2}: $\gamma > 1/2$.
    We know that $W$ is convex and $s \longmapsto (1+s \chi_i)\mu$ is affine. Therefore, their composition $\Phi$ is convex.
    Using Fact~\ref{fact_convex_upper_bound}, $\Phi(\gamma)+ \Phi(-\gamma) \le \Phi(1) + \Phi(-1)$.

    Let $\mu^0_i(x) = \mu(x)\mathbf{1}_{\{x_i=0\}}$, $\mu^1_i(x) = \mu(x)\mathbf{1}_{\{x_i=1\}}$. Thus, $(1+ \chi_i) \mu = 2 \mu^1_i$ and $(1- \chi_i) \mu = 2 \mu^0_i$. By positive homogeneity of $W$, we get 
    \[\Phi(1) =W(2 \mu_i^1) = 2W( \mu_i^1) , \quad  \Phi(-1)=W(2 \mu_i^0) =2W(\mu_i^0) .\]

    Therefore, 
    \[ \frac{\Phi(\gamma)+ \Phi(-\gamma)}{2} \le W( \mu_i^0)+ W( \mu_i^1). \]

    Additionally, we have $q_i(\mu) =W( \mu_i^0)+ W( \mu_i^1) - \alpha m(\mu) \le W(\mu)$ where the inequality follows from Lemma~\ref{lem_basic_potential_properties}. Substituting $m(\mu)=1$ and $W(\mu)=\Phi(0)$, we get $\Delta_{\gamma} \le \alpha.$
    Since $\gamma > 1/2$ and $N \ge 2$, we obtain
    \[\Delta_{\gamma} \le \alpha \le  6 \alpha (\log N) \gamma^2.\]

    Taking $C_b' = \max\{\frac{\widehat{C}_b}{2},6\}$ completes the proof.
\end{proof}

\begin{corollary}[Total Potential Increase]\label{corollary_total_potential}
    Fix a probability distribution $\mu_0$. Suppose $\mathcal{A}$ is a finite noisy randomized decision tree run on an input $X \sim \mu_0$, with stopping time $T$ and let $\gamma_t \in [0,1]$ be the strength of its $t$-th noisy query. If $\mu_t$ is the posterior distribution after $t$ queries, then,
    \[\E[W(\mu_T)] - W(\mu_0) \le C_b' \alpha \log{N} \E\left[\sum_{t=1}^T \gamma_t^2\right].\]
\end{corollary}
\begin{proof}
    Let $\mathsf{H}_{t-1}$ denote the complete transcript before $t$. Conditioned on this transcript, the current posterior and query strength are fixed. Let $\mu_t = \mu_T$ and $\gamma_t=0$ for all $t > T$. At $t>T$, since the posterior remains unchanged, the conditional expectation increase is zero and the query cost is also zero.
    
    From previous Proposition~\ref{prop_one_potential_increase}, we have for $t\ge 1$,
    \[\E[W(\mu_t)| \mathsf{H_{t-1}}] - W(\mu_{t-1}) \le C_b' \alpha \log{N} \gamma_t^2.\]

    Taking expectation and using tower property gives 
    \[\E[W(\mu_t)] - \E[W(\mu_{t-1})]  \le C_b' \alpha \log{N} \E[\gamma_t^2].\]

    Summing over $t \ge 1$ generates a telescoping series on left-hand side and gives us 
    \[ \E[W(\mu_T)] - \E[W(\mu_{0})]  \le C_b' \alpha \log{N} \sum_{t\ge 1} \E[\gamma_t^2]. \]
     By padding and linearity we get $\sum_{t\ge 1} \E[\gamma_t^2] = \E[\sum_{t\ge 1} \gamma_t^2] = \E\left[\sum_{t= 1}^T \gamma_t^2\right]$. Since the tree is finite, these sums can have only finitely many potentially nonzero terms. We then obtain,
    \[\E[W(\mu_T)] - \E[W(\mu_{0})]  \le C_b' \alpha \log{N} \E\left[\sum_{t= 1}^T \gamma_t^2\right].\]
    Since $\mu_0$ is a fixed distribution, $\E[W(\mu_{0})] = W(\mu_0)$.
\end{proof}

\subsection{Initial and terminal potential}

\begin{lemma}[Initial potential]\label{lem_initial_potential}

    Suppose a probability distribution $\mu$ satisfies,
    \begin{equation}\label{eq_initial_potential_hypothesis}
       \E[\beta^2_{\mathsf{H}_{T}}] \le \frac{\alpha}{2}\E[Q_{\mathcal{B}}], 
    \end{equation}
    for every auxiliary forecasting algorithm $\mathcal{B}$ with query count $Q_{\mathcal{B}}$ and terminal transcript $\mathsf{H}_{T}$ where $T=Q_{\mathcal{B}}$.
The initial potential $W$ then satisfies 
    \[ W(\mu) \le \frac{1}{100} .\]
\end{lemma}

\begin{proof}
    Fix an auxiliary forecasting algorithm $\mathcal{B}$. For each terminal transcript $h$ of positive probability, let $z_h \in [0,1]$ be the forecast determined by $\mathsf{H}_{T}=h$. Using the definition of $W$, we want to bound the quantity $\sum_{x \in D} \mu(x) L_{\mathcal{B}}(x)$ which is exactly the expectation of $L_{\mathcal{B}}(X)$ over $X \sim \mu$.
    
    Define $\theta_h = \Pr[f(X)=1 \mid  \mathsf{H}_{T}=h]$. Then the expected score is 
    \[ \E[\operatorname{score}_{Br}(f(X),Z) \mid \mathsf{H}_{T}=h] = \theta_h (1-4(1-z_h)^2)+ (1- \theta_h)(1-4z_h^2) \le (2 \theta_h -1)^2.   \]

    The last inequality comes from the fact that maximum of this expression occurs at $z_h = \theta_h$.
    Let $\beta_{h}$ be the posterior bias and thus, $\beta_{h} = 2 \theta_h-1$. Taking expectation over all the transcripts and using Equation~\ref{eq_initial_potential_hypothesis}, we get,
    \[ \E[\operatorname{score}_{Br}(f(X),Z)]  \le \E[\beta_{\mathsf{H}_{T}}^2] \le \frac{\alpha}{2}\E[Q_{\mathcal{B}}].   \]

    Let $p_t$ be the action distribution at the visited history immediately before decision $t$. The distribution $p_t$ is specified by the history of previous actions and answers, before sampling the next action, rather than by conditioning on all presampled internal randomness.
     Using Fact~\ref{fact_entropy_bound} and $\tau = \frac{\alpha}{b \log N} $, we get 
    \[ \tau \E\left[\sum_{t=0}^{Q_{\mathcal{B}}}\Ent(p_t) \right]  \le \tau \log N (\E[Q_{\mathcal{B}}]+1) = \frac{\alpha}{b}(\E[Q_{\mathcal{B}}]+1).\]

    Finally, we get 
    \[ \sum_x \mu(x)L_{\mathcal{B}}(x) \le \frac{\alpha}{2} \E[Q_{\mathcal{B}}] - \alpha \E[Q_{\mathcal{B}}] + \frac{\alpha}{b}(\E[Q_{\mathcal{B}}]+1). \]
    Since $b \ge 100$ and $\alpha \le 1$, taking supremum over all such auxiliary forecasting algorithms gives us,
    \[ W(\mu) \le \frac{\alpha}{b} \le \frac{1}{100}. \]
\end{proof}

\begin{lemma}[Terminal potential]\label{lem_terminal_potential}
    Let $\mathcal{A}$ be a noisy oracle randomized algorithm computing $f$ within error at most $1/3$ on every input, with stopping time $T$. Let $\mathsf{H}_T$ be its complete terminal transcript and $\mu_T(x) :=\Pr[X=x \mid \mathsf{H}_T] $ be the terminal posterior distribution under any input distribution. Then the terminal potential $W(\mu_T)$ satisfies 
    \[ \E[W(\mu_T)] \ge \frac{1}{9} .\]
\end{lemma}
\begin{proof}
    
    Let $\widehat{Y} \in \{0,1\}$ be the output of the noisy algorithm $\mathcal{A}$ on input $X \sim \mu$ for any probability distribution $\mu$. $\mathsf{H}_T$ determines $\widehat{Y}$. Similarly, terminal posterior bias is $\beta_T =2\Pr[f(X)=1 \mid \mathsf{H}_T] -1$.
    By Lemma~\ref{lem_basic_potential_properties}, we have $W(\mu_T) \ge S(\mu_T) = \beta_T^2$. 

    \[ \E\bigl[ (2 \widehat{Y}-1)(2 f(X)-1) \mid \mathsf{H}_T \bigr] = (2 \widehat{Y}-1) \E\bigl[ 2 f(X)-1 \mid \mathsf{H}_T \bigr] =(2 \widehat{Y}-1) \beta_T \le |\beta_T|. \]

    We get the first equality because $\widehat{Y}$ is determined by $\mathsf{H}_T$. 
    Now, taking expectation over all the terminal transcripts, we get
    \[ \E\bigl[ (2 \widehat{Y}-1)(2 f(X)-1) \bigr]  \le \E |\beta_T|. \]

    But $(2 \widehat{Y}-1)(2 f(X)-1) = 1- 2 \mathbf{1}_{\{\widehat{Y} \ne f(X)\}}$ and therefore, $\E\bigl[ (2 \widehat{Y}-1)(2 f(X)-1) \bigr] = 1- 2\Pr[\widehat{Y}\ne f(X)]$.
    Furthermore, $\Pr[\widehat{Y}\ne f(X)] \le 1/3$ since $\mathcal{A}$ computes $f(X)$ with error at most $1/3$. Hence, $\E |\beta_T| \ge 1/3$.

    By Cauchy-Schwarz, we get $(\E |\beta_T|)^2 \le  \E(\beta_T^2)$ and finally we get 
    \[ 1/9 \le (\E |\beta_T|)^2 \le  \E(\beta_T^2) \le \E W(\mu_T). \]
\end{proof}

\subsection{Completing the Proof of Main Theorem}

We now prove the main result of the paper. 

\begin{theorem}\label{thm_noisy_randomized}
    For a non-constant partial Boolean function $f: D \subseteq \{0,1\}^n \rightarrow \{0,1\}$, 
    \[ \R(f) \le C \log{(n+1)}\noisyR(f). \]
\end{theorem}
\begin{proof}
    If $\R(f) \le 2c$, then the result follows from $\noisyR(f) = \Omega(1)$ from Lemma~\ref{lem_noisy_constant_lower}. Otherwise, we can take $\alpha = 2c/\R(f)$ and the distribution $\mu$ from Lemma~\ref{lem_hard_dist_existence}. Since $c\ge 1$ and $2c <\R(f) \le n$, we have $ 1/n \le \alpha \le 1$. Moreover, for every auxiliary forecasting algorithm $\mathcal{B}$ with terminal transcript $\mathsf{H}$ and query cost $Q_{\mathcal{B}}$, 
      \[\E[\beta_{\mathsf{H}}^2] \le \frac{c}{\R(f)} \E[Q_{\mathcal{B}}] = \frac{\alpha}{2}\E[Q_{\mathcal{B}}].\]
      We can therefore apply Lemma~\ref{lem_initial_potential}.
     Let $\mathcal{A}$ be a finite noisy randomized algorithm computing $f$ within error of $1/3$ with stopping time $T$. Applying Lemma~\ref{lem_initial_potential}, Lemma~\ref{lem_terminal_potential} and Corollary~\ref{corollary_total_potential}, we get
    \[ \frac{1}{9} - \frac{1}{100} \le \E[W(\mu_T)] - W(\mu) \le C_b' \alpha \log{N} \cost(\mathcal{A},\mu), \]
    where $\cost(\mathcal{A},\mu) = \E_{X \sim \mu}[\cost(\mathcal{A},X)]$.
    Resubstituting $N=n+1$, $\alpha = 2 c /\R(f)$ , we get
    \[ \cost(\mathcal{A},\mu) \ge \frac{91}{1800 c C_b' \log{(n+1)} } \R(f).   \]

    We know that $\max_{x \in D} \cost(\mathcal{A},x) \ge \cost(\mathcal{A},\mu)$. Since $\mathcal{A}$ is arbitrary, taking infimum over all such algorithms, we get
    \[ \noisyR(f) =  \inf_{\mathcal{A}} \max_{x \in D} \cost(\mathcal{A},x) = \Omega\left(\frac{\R(f)}{\log{(n+1)}}\right). \]
\end{proof}

\begin{corollary}\label{cor_randomized_composition}
    For non-constant Boolean functions $f: D \subseteq\{0,1\}^n \rightarrow \{0,1\}$ and $g:D' \subseteq\{0,1\}^m \rightarrow \{0,1\}$,
    \[ \R(f \circ g) = \Omega\left(\frac{\R(f)\R(g)}{\log{(n+1)}}\right).\]
\end{corollary}
\begin{proof}
    The result follows immediately from Theorem~\ref{thm_noisy_randomized} and Lemma~\ref{lem_noisy_lower_bound}.
\end{proof}

\newpage
\printbibliography
\appendix

\section{Intermediate Estimates for Softmax Second-Order Bound}\label{appendix_softmax_bound}

\begin{proposition*}[Restatement of Proposition~\ref{prop_curvature_bound}]
     For every probability distribution $\mu$ and every coordinate $i \in [n]$,
    \begin{equation}
        \left. \frac{d^2}{ds^2} W((1+s \chi_i)\mu) \right|_{s=0} \le C_b \alpha \log{N},
    \end{equation}
    where $C_b$ is a constant depending on $b$.
\end{proposition*}
\begin{proof}
Set $K = 36b \log{N}, \; \varepsilon = \frac{\alpha}{N^4}$. Fix a coordinate $i \in [n]$. For a non-negative measure  $\xi$, write 
\[ \Phi_{\xi}(s) = W((1+s \chi_i(x))\xi), \quad
 D_{\xi}^+ = \Phi_{\xi}(1) - \Phi_{\xi}(0) - \Phi_{\xi}'(0), \qquad D_{\xi}^- = \Phi_{\xi}(-1) - \Phi_{\xi}(0) + \Phi_{\xi}'(0).\]

Let $u(\xi)$ be the number of coordinates that are non-constant on $\operatorname{supp}(\xi)$. Equivalently,
\[ u(\xi) = |\{j \in [n] \mid \exists \; x,x' \in \operatorname{supp}(\xi)\;  \text{such that} \; x_j\ne x_j'  \}|, \]
and $u(0)=0$.
We will first prove the following using induction on $u(\xi)$:
\begin{equation}\label{eq_strong_second_order}
\Phi_{\xi}''(0) -KH(D^+_{\xi},D^-_{\xi}) \le 2u(\xi)\varepsilon m(\xi).    
\end{equation}

\textbf{Base Case.}

For $\xi = 0$, above inequality is immediate. When $\xi \ne 0$ and $u(\xi)=0$, this indicates that $\operatorname{supp}(\xi)$ is non-empty but contains exactly one input $x^*$. Let $2x_i^*-1 = \sigma$. Then, by positive homogeneity of $W$, we get $W((1+ \sigma s)\xi) = (1+ \sigma s)W(\xi)$. Since this is affine in $s$, $\Phi_{\xi}''(0) =0$ and both $D^+_{\xi},D^-_{\xi}$ are $0$. This covers the base case.

\textbf{Induction Step.}
Now, fix an integer $k \ge 1$ and suppose that the Equation~\ref{eq_strong_second_order} holds for every non-negative measure $\eta$ with $u(\eta) \le k-1$ undetermined coordinates. Suppose $u(\xi) = k$. Let us first look at the case when $i$ is constant on $\operatorname{supp}(\xi)$. By the same argument used in the base case, $\Phi_{\xi}$ is affine in $s$ and the claim follows. So let us look at the case when $i$ is not constant and hence undetermined.

Since $\xi \ne 0$, we can normalize it to be a probability distribution $\nu = \frac{\xi}{m(\xi)}$ and since normalization does not change the support, we get $u(\xi) = u(\nu)=k$. Using positive homogeneity we get
\[ \Phi_{\xi}(s) = m(\xi) \Phi_{\nu}(s), \quad D_{\xi}^+ = m(\xi)D_{\nu}^+, \quad D_{\xi}^- = m(\xi)D_{\nu}^-. \]

Moreover, $H$ is also positively homogeneous. Therefore, we can just prove the bound in Equation~\ref{eq_strong_second_order} for $\nu$. Henceforth, we will suppress the subscript for $\nu$ in $\Phi_{\nu},D_{\nu}^+,D_{\nu}^-$.

Now, set $R = \Phi''(0) - KH(D^+,D^-)$ so that it suffices to prove $R \le 2k\varepsilon$. Applying Lemma~\ref{lem_softmax_bound} gives 
\begin{equation}
    R \le \E_{a \sim \pi} \left[q_a''(0) - KH(d_a^+,d_a^-)\right] + \varepsilon.
\end{equation}

Next, we bound the contribution of each possible first action.
Let $\mathcal{J} \subseteq[n]$ be the set of coordinates determined on $\operatorname{supp}(\nu)$.
Let $\nu_s = (1+s \chi_i) \nu$.
\begin{itemize}
    \item \noindent\emph{Stop action.}

Let $\beta = \E_\nu[2 f(X)-1], \qquad \zeta = \E_{\nu}[(2f(X)-1)\chi_i(X)], \qquad r= \E_{\nu}[\chi_i(X)]$ . 

We have $q_*(s) = S(\nu_s)$. The mass of $\nu_s$ is $m(\nu_s) = 1+rs$ and $\sum_x \nu_s(x)(2f(x)-1) = \beta+ s \zeta$. 
Therefore, $q_*(s) = \frac{(\beta+ s \zeta)^2}{1+rs}$ and thus, $q_*''(0) = 2(\zeta-r \beta)^2$. We also get
\[ d^+_* = q_*(1)-q_*(0)- q_*'(0) = \frac{(\zeta-r \beta)^2}{1+r}, \quad  d^-_* = q_*(-1)-q_*(0)+ q_*'(0) = \frac{(\zeta-r \beta)^2}{1-r}.\]
Because coordinate $i$ is undetermined, $r \in (-1,1)$ and hence, $d_*^+,d_*^- \ge 0$.

Substituting in $H$ gives us $H(d_*^+,d_*^-) = \frac{(\zeta-r\beta)^2}{2}$.
We finally get $q_*''(0) = 4H(d_*^+,d_*^-)$ and since $K \ge 4$, $q_*''(0) - KH(d_*^+,d_*^-) \le 0$.
\item \noindent\emph{Querying an undetermined coordinate.}

Suppose action $j$ queries a coordinate $j \notin \mathcal{J}$ that is undetermined under $\nu$.
Define the child measures $\nu^1_j(x) = \nu(x)\mathsf{1}_{x_j=1}$ and $\nu^0_j(x) = \nu(x)\mathsf{1}_{x_j=0}$. Clearly, coordinate $j$ is determined under these child measures. Therefore, $u(\nu^1_j) \le k-1$ and $u(\nu^0_j) \le k-1$. Thus, the induction hypothesis applies to $\nu^1_j$ and $\nu^0_j$.

\[\Phi_{\nu^1_j}''(0) -KH(D^+_{\nu^1_j},D^-_{\nu^1_j}) \le 2u(\nu^1_j)\varepsilon m(\nu^1_j), \quad \Phi_{\nu^0_j}''(0) -KH(D^+_{\nu^0_j},D^-_{\nu^0_j}) \le 2u(\nu^0_j)\varepsilon m(\nu^0_j).\]

Furthermore, using $m(\nu^0_j)+ m(\nu^1_j) = m(\nu)=1$ and $u(\nu^1_j) \le k-1$ and $u(\nu^0_j) \le k-1$, we get,
\begin{equation}\label{eq_intermediate_unqueried}
\Phi_{\nu^1_j}''(0) + \Phi_{\nu^0_j}''(0)  \le KH(D^+_{\nu^1_j} ,D^-_{\nu^1_j}) + KH(D^+_{\nu^0_j},D^-_{\nu^0_j}) + 2(k-1)\varepsilon .    
\end{equation}
Now, we have 
\[ q_j(s)= -\alpha m(\nu_s)+ \Phi_{\nu^1_j}(s)+ \Phi_{\nu^0_j}(s), \quad d^+_j = q_j(1)-q_j(0)-q_j'(0), \quad d^-_j = q_j(-1)-q_j(0)+q_j'(0). \]
Since $m$ is affine in $s$, we get $ q_j''(s)=  \Phi''_{\nu^1_j}(s)+ \Phi''_{\nu^0_j}(s)$.
Therefore, substituting $q_j$ in the definition of $d_j^+,d_j^-$, we get $d^+_j = D^+_{\nu^0_j}+D_{\nu^1_j}^+,\;  d^-_j = D^-_{\nu^0_j}+D_{\nu^1_j}^-$.
Using super-additivity of $H$, we get $H(D^+_{\nu^1_j},D^-_{\nu^1_j})+H(D^+_{\nu^0_j},D^-_{\nu^0_j}) \le H(D^+_{\nu^1_j}+D^+_{\nu^0_j},D^-_{\nu^1_j}+D^-_{\nu^0_j}) = H(d^+_j,d^-_j)$.

Substituting these in Equation~\ref{eq_intermediate_unqueried} gives us
\[ q_j''(0)-KH(d^+_j,d^-_j) \le 2(k-1) \varepsilon.\]

\item \emph{Querying a determined coordinate}

Fix $j \in \mathcal{J}$.
Suppose that coordinate $j$ is determined on $\operatorname{supp}(\nu)$. Therefore, from the determined query identity in Equation~\ref{eq_determined_coordinate_W}, we have 
\[q_j(s) = \Phi(s)- \alpha m(s).\]

Because $m(s) = m(\nu_s) = 1+rs$, where $r = \E_{\nu}[\chi_i(X)]$ and so it is affine. $m'(s)=r$ and $m''(s)=0$. We have $q_j(s) = \Phi(s)- \alpha(1+rs)$. Hence, $q_j'(0) = \Phi'(0)-\alpha r, \quad q_j''(0)= \Phi''(0)$.  
\[d_j^+ = q_j(1) - q_j(0)- q_j'(0) = (\Phi(1)- \alpha(1+r)) - (\Phi(0)-\alpha) - (\Phi'(0))+ \alpha r = D^+.\]

Similarly, we get $d_j^- = D^-$. Therefore, finally we get 
\[ q_j''(0)- KH(d_j^+,d_j^-) = \Phi''(0)-KH(D^+,D^-) = R. \]
Furthermore, $m(0)=1$ and $q_j(0) = \Phi(0)- \alpha$. Therefore,
\[ \pi_j = \exp\left( \frac{q_j(0)- \Phi(0)}{\tau} \right) = e^{-\alpha/ \tau} = N^{-b}. \]

Let $p_{\text{det}} = \sum_{j \in \mathcal{J}} \pi_j$. Since there are at most $n$ determined coordinates, $p_{\text{det}} \le n N^{-b} \le 1/2$. 
\end{itemize}
\emph{Combining the three cases.}
The stop action contributes at most $0$, every query of an undetermined coordinate contributes at most $2(k-1) \varepsilon$ and every query of a determined coordinate contributes $R$. Therefore, $R \le 2(k-1)\varepsilon \sum_{j \in [n]\backslash \mathcal{J}}\pi_j + p_{\text{det}}R + \varepsilon.$  Since $\pi_*+ \sum_{j \in [n]\backslash \mathcal{J}}\pi_j  + \sum_{j \in \mathcal{J} }\pi_j=1$, we get
\[ R \le 2(k-1)\varepsilon(1-p_{\text{det}})+ p_{\text{det}}R + \varepsilon. \]

Rearranging and using $p_{\text{det}} \le 1/2$, we get $R \le 2(k-1)\varepsilon+2 \varepsilon = 2k\varepsilon$. Further multiplying by $m(\xi)$ completes the induction. 

For a probability distribution $\mu$, using positive homogeneity of $W$ and $q_i(\mu) \le W(\mu)$, we get $D^+_{\mu}+D^-_{\mu} = 2W(\mu_i^1)+ 2W(\mu_i^0)-2W(\mu) \le 2 \alpha $ and hence $H(D^+_{\mu}, D^-_{\mu}) \le \alpha/2$. Using the inductive bound and the fact that $u(\mu)\le n$, we obtain
\[ \Phi''(0) \le \frac{K \alpha}{2}+2n\varepsilon  \le C_b \alpha \log{N}.\]

\end{proof}

\begin{lemma*}[Restatement of Lemma~\ref{lem_softmax_bound}]
Fix a coordinate $i$ and a distribution $\nu$ and assume $i$ is undetermined on $\operatorname{supp}(\nu)$. For $s \in [-1,1]$, write $\Phi(s)= W\left( (1+ \chi_i(x) s)\nu \right)$. 

   For $K= 36 b \log{N}$ and $\varepsilon = \frac{\alpha}{N^4}$,
    \[ \Phi''(0) \le \E[q_a''(0)]+K\left( H(D^+,D^-) - \E[H(d_a^+,d_a^-)] \right) +\varepsilon. \]
\end{lemma*}
\begin{proof}
    
    For $s \in [-1,1]$, define the softmax action probabilities 
    \[ \pi_a(s) = \exp\left( \frac{q_a(s)-\Phi(s)}{\tau m(s)} \right). \]
    All sums and expectations in this proof are over the full action set $\ACT$ and the expectations are taken under $\pi_a = \pi_a(0)$.
    We write $\pi_a = \pi_a(0)$. We have $\sum_a \pi_a(s) = 1$ for every $s$ from Equation~\ref{eq_exp_sum_W}. We also write $r = \E_{\nu}[\chi_i(X)]$ and $m(s)=1+rs$. Define 
    \[ \delta_a =\Phi(0)-q_a(0), \quad w_a = \tau \frac{\pi_a'(0)}{\pi_a}. \]
    Differentiating the equation for $\log{\pi_a(s)}$ at $s=0$ gives us $w_a = q_a'(0)-\Phi'(0)+r\delta_a$.

    Furthermore, taking the first order differentiation of $\sum_a \pi_a(s) = 1$  gives us $\E[w_a] = \sum_a w_a \pi_a =0$.
    
    Rewriting the definition of $\pi_a(s)$ as $q_a(s) = \Phi(s) + \tau m(s) \log{\pi_a (s)}$,  taking second derivative and using $m(0)=1$ and $m''(s)=0$ gives us
    \[q_a''(0) = \Phi''(0)+ 2 \tau r \frac{\pi_a'(0)}{\pi_a}+ \tau \frac{\pi_a''(0)}{\pi_a} - \frac{w_a^2}{\tau}.\]
    Multiplying by $\pi_a$, taking summation over all $a$ and using the fact that $\sum_a \pi_a'(0)=0$ and $\sum_a \pi_a''(0)=0$, we get 
    \begin{equation}\label{eq_intermediate_second_moment}
        \Phi''(0)  = \E[q_a''(0)]+\frac{\E w_a^2}{\tau}.    
    \end{equation}
    Therefore, it suffices to bound the term $\E w_a^2$.

   Define 
   \[ v_a^+ = \tau m(1) \log{\frac{\pi_a(1)}{\pi_a}}, \quad v_a^- = \tau m(-1) \log{\frac{\pi_a(-1)}{\pi_a}}. \]
    Moreover,
    \begin{equation}\label{eq_v_a_expectation}
     \E\left[\exp\left(\frac{v_a^+}{\tau m(1)}\right)\right] = \sum_{a} \pi_a \frac{\pi_a(1)}{\pi_a}=1, \quad \E\left[\exp\left(\frac{v_a^-}{\tau m(-1)}\right)\right] = \sum_{a} \pi_a \frac{\pi_a(-1)}{\pi_a}=1.      
    \end{equation}

   Applying Jensen's inequality to above equations gives us $\E[v_a^+] \le 0$ and $\E[v_a^-] \le 0$.
   Using the definitions of $w_a, d_a^+,d_a^-,D^+,D^-$, we obtain
   \[v_a^+ = d_a^+ + w_a- D^+ , \quad v_a^- = d_a^- - w_a- D^-.\]
    Taking expectations in the preceding equation and using $\E[w_a]=0$, we get 
    \begin{equation}\label{eq_v_a_d_a_D}
     \E[v_a^+] = \E[d_a^+]-D^+ , \quad \E[v_a^-] = \E[d_a^-]-D^-.     
    \end{equation}

   Since, $\E[v_a^+] \le 0$ and $\E[v_a^-] \le 0$, $D^+-\E[d_a^+] \ge 0$ and $D^--\E[d_a^-] \ge 0$.

   We now classify the actions as regular or exceptional just for the purpose of this proof. If an action has $\delta_a \le \alpha$, then it is called regular, otherwise it is called exceptional. Note that for all query actions besides stopping, we have $\Phi(0)- \alpha \le q_a(0)$ and hence it is regular. Only the stopping action can be exceptional, namely when $\delta_* > \alpha$. Let $p$ be the total probability of exceptional actions. Then
   \[ p = \sum_{a: \delta_a > \alpha} \pi_a \le e^{-\alpha/\tau} = N^{-b}. \]

   Now, first note since $m(0)=1$, we have $\pi_a = e^{-\delta_a/\tau}$ and $\pi_a(1)\le 1 $ and $\pi_a(-1) \le 1$. Using the definition of $v_a^+,v_a^-$ and monotonicity of logarithm and $m(0)=1$, we get $v^+_a \le m(1)\delta_a$ and $v^-_a \le m(-1)\delta_a$.

   We know that $q_i(\nu) \le \Phi(0)$ and $q_i(\nu) = -\alpha +(\Phi(1)+\Phi(-1))/2$, we get $D^+ + D^- \le 2 \alpha$. Convexity of $q_a(s)$ gives $d_a^+,d_a^- \ge 0$.
   Using $D^+ + D^- \le 2 \alpha$ and  $d_a^+,d_a^- \ge 0$ gives 
   \[ v^+_a + v^-_a =d_a^+ + d_a^- -D^+ - D^- \ge -2 \alpha. \]

   Let us now look at a regular action. Since $\delta_a \le \alpha$ and $m(1),m(-1) \le 2$, we get $v^+_a \le m(1)\delta_a \le 2 \alpha$ and $v^-_a \le m(-1)\delta_a \le 2 \alpha$ and combining these with $v^+_a + v^-_a \ge  -2 \alpha$ gives us $v^+_a,v^-_a \ge -4 \alpha$. Thus, $|v_a^+| \le 4 \alpha$ and $|v_a^-| \le 4 \alpha$. Moreover, since $d_a^+ + d_a^-= v_a^+ + v_a^- + D^+ +D^-$, we get
   \[ d_a^+ + d_a^- \le (m(1)+m(-1))\delta_a + (D^++ D^-) \le 2\alpha + 2 \alpha = 4 \alpha. \]
   We also need a global bound that holds for every action, regular or exceptional.

    \[ \delta_j \le \alpha, \quad \delta_* = W(\nu)-S(\nu)\le 1+ \frac{\alpha}{b}. \]
    Thus, $\delta_a \le 3$ for every action. Using earlier inequalities, we get $v_a^{+}\le 6$, $v_a^{-}\le 6$, $v_a^{+}\ge -8$, $v_a^{-}\ge -8$ and $d_a^++d_a^- \le 2 \delta_a+2 \alpha \le 8$. Therefore, we will use $8$ as the rough global bound for every action. 

    Let $0 < \eta\le 1$. We know that the map $ y \longmapsto y ^{\eta}$ is concave when $y>0$. Now, $\tau m(1) \le 2 \tau \le 4 \alpha$, taking $\eta = \frac{\tau m(1)}{4 \alpha}$, and applying Jensen's inequality, we get
    \[  \E [e^{v_a^+/{4 \alpha}}] = \E [(e^{v_a^+/{4 \eta \alpha}})^{\eta}] \le \left(\E[(e^{v_a^+/{4 \eta \alpha}})] \right)^{\eta}.  \]
    Using Equation~\ref{eq_v_a_expectation}, we get $\E [e^{v_a^+/{4 \alpha}}] \le 1$. Similar calculation for $v_a^-$ gives $\E [e^{v_a^-/{4 \alpha}}] \le1$.

    Substituting $y = v_a^+/{4 \alpha}$ in $e^{y} \ge 1+y+\frac{y^2}{3} \mathbf{1}_{\{|y|\le 1\}}$, we get 
    \[ e^{v_a^+/{4 \alpha}} \ge 1+ (v_a^+/{4 \alpha}) +\frac{(v_a^+/{4 \alpha})^2 \mathbf{1}_{\{|v_a^+|\le 4 \alpha\}}}{3}. \]

    Taking expectation and using the bound $\E[e^{v_a^+/{4 \alpha}}] \le 1$, and after re-arranging, we get  
    \[\E[(v_a^+)^2\mathbf{1}_{\{|v_a^+|\le 4 \alpha\}}] \le 12 \alpha(-\E(v_a^+)).\]
    Note that only for exceptional actions, we can get $|v_a^+| > 4 \alpha$ or $|v_a^-| > 4 \alpha$ and for those actions, since the total probability is at most $p$, together with $|v_a^+|\le 8$ and $|v_a^-|\le 8$, these actions contribute at most $64p$.
    Using Equation~\ref{eq_v_a_d_a_D}, and similar calculation for $v^-_a$ we finally get, 
    \begin{equation}\label{eq_second_moment_v_a}
     \E[(v_a^+)^2] \le 12 \alpha (D^+ - \E[d_a^+])+ 64p, \quad \E[(v_a^-)^2] \le 12 \alpha (D^- - \E[d_a^-])+64p.   
    \end{equation}
    
    We now combine the two endpoints. Define $z$ as 
    \[  z = \begin{cases}
        \frac{D^-}{D^+ + D^-}, & D^++D^- > 0, \\
        \frac{1}{2}, & D^++D^- = 0.
    \end{cases} \]

    Note that $z D^+ = (1-z)D^-$ and $H(D^+,D^-) = z^2D^++(1-z)^2D^-$.

    For each action $a$, define 
    \[ A_a =z d_a^+ - (1-z)d_a^- , \quad B_a =z^2d_a^++(1-z)^2d_a^- - H(d_a^+,d_a^-).  \]
    Since $H(u,v)$ can also be written as $H(u,v) = \min_z \{z^2u+(1-z)^2v\}$, $B_a \ge 0$ and direct expansion gives $A_a^2 = (d_a^++d_a^-)B_a$.

    Since on a regular action $d_a^++d_a^- \le 4 \alpha$, therefore $A_a^2 \le 4 \alpha B_a$. For every action, the rough bound gives $d_a^++d_a^- \le 8$ and therefore $|A_a| = |(z)d_a^+-(1-z)d_a^-| \le |d_a^++d_a^-|\le 8$ and thus $|A_a|^2 \le 64$. Finally, we get 
    \begin{equation}\label{eq_A_a_B_a}
        \E[A_a^2] = \E[A_a^2 \mathbf{1}_{\textbf{regular}}]+ \E[A_a^2 \mathbf{1}_{\textbf{exceptional}}] \le 4 \alpha\E[ B_a \mathbf{1}_{\textbf{regular}}]+ 64p \le 4 \alpha\E[ B_a]+ 64p, 
    \end{equation}
    where the last inequality uses the fact that $B_a \ge 0$.
    Now, define $G = H(D^+,D^-)-\E[H(d_a^+,d_a^-)]$ which can be decomposed into three non-negative terms as follows
    \[ G = z^2(D^+- \E[d_a^+])+(1-z)^2(D^- - \E[d_a^-])+ \E[B_a]. \]
    \[ w_a = z v^+_a-(1-z)v^-_a - A_a. \]
    Using $(x+y+z)^2 \le 3(x^2+y^2+z^2)$, Equation~\ref{eq_second_moment_v_a}, Equation~\ref{eq_A_a_B_a}, we get
    \[\E[w_a^2] \le 3z^2(\E[(v_a^+)^2])+3(1-z)^2(\E[(v_a^-)^2])+3\E[A_a^2] \le 36\alpha G+384p. \]
        Using $\alpha \ge 1/n \ge 1/N$ and $N\ge 2$ and $b \ge 100$, we get $\frac{384p}{\tau \varepsilon}\le 384 b (\log N)N^{6-b} <1$.
    Dividing by $\tau = \alpha/(b \log N)$ and using $p \le N^{-b}$, we get 
    \[ \frac{\E[w_a^2]}{\tau} \le 36b \log{N}\left( H(D^+,D^-) - \E[H(d_a^+,d_a^-)] \right)+ \frac{\alpha}{N^4}. \]
    
    Resubstituting this in Equation~\ref{eq_intermediate_second_moment} gives us the required bound.
\end{proof}

\end{document}